\documentclass[a4paper,fleqn]{cas-dc}

\newcommand{\githublink}{\href{https://github.com/HongkunSun/ParaTransCNN}{https://github.com/HongkunSun/ParaTransCNN}}
\hypersetup{
	colorlinks=true,
	linkcolor=blue,
	urlcolor=magenta,
}
\usepackage[numbers]{natbib}

\def\tsc#1{\csdef{#1}{\textsc{\lowercase{#1}}\xspace}}
\tsc{WGM}
\tsc{QE}

\begin{document}
\let\WriteBookmarks\relax
\def\floatpagepagefraction{1}
\def\textpagefraction{.001}

\shorttitle{ParaTransCNN: Parallelized TransCNN Encoder for Medical Image Segmentation}     

\shortauthors{H. Sun et al.}  

\title [mode = title]{ParaTransCNN: Parallelized TransCNN Encoder for Medical Image Segmentation}  

\tnotemark[<tnote number>] 


%
\author[1]{Hongkun Sun}
\ead{sunhongkun888@163.com}
\author[1]{Jing Xu}
\ead{jingxu@amss.ac.cn}
\author[2]{Yuping Duan\corref{cor1}}
\ead{doveduan@gmail.com}
\cortext[cor1]{Corresponding author}
\address[1]{School of Statistics and Mathematics, Zhejiang Gongshang University, Hangzhou, 310018, China}
\address[2]{School of Mathematical Sciences, Beijing Normal University, Beijing, 100875, China}

















\begin{abstract}
The convolutional neural network-based methods have become more and more popular for medical image segmentation due to their outstanding performance. However, they struggle with capturing long-range dependencies, which are essential for accurately modeling global contextual correlations. Thanks to the ability to model long-range dependencies by expanding the receptive field, the transformer-based methods have gained prominence. Inspired by this, we propose an advanced 2D feature extraction method by combining the convolutional neural network and Transformer architectures. More specifically, we introduce a parallelized encoder structure, where one branch uses ResNet to extract local information from images, while the other branch uses Transformer to extract global information. Furthermore, we integrate pyramid structures into the Transformer to extract global information at varying resolutions, especially in intensive prediction tasks. To efficiently utilize the different information in the parallelized encoder at the decoder stage, we use a channel attention module to merge the features of the encoder and propagate them through skip connections and bottlenecks. Intensive numerical experiments are performed on both aortic vessel tree, cardiac, and multi-organ datasets. By comparing with state-of-the-art medical image segmentation methods, our method is shown with better segmentation accuracy, especially on small organs. The code is publicly available on \githublink.
\end{abstract}


\maketitle

\section{Introduction}
Manual marking of the different characteristics of lesion and non-lesion areas in medical images is both time-consuming and challenging for doctors. Therefore, Computer-Aided Diagnosis (CAD) \cite{CAD_cnn}, \cite{CAD_transformers} systems can accurately segment the lesion area and improve the effectiveness of screening and diagnosing diseases by assisting doctors in their decision-making process.

The U-Net \cite{u_net}, a groundbreaking architecture for medical image segmentation, has achieved considerable success in various segmentation tasks. It learns to feature information from input images through downsampling in the encoder, combines low-level and high-level semantic information through skip connections, and restores detailed image information through upsampling to generate segmentation masks. Since then, the symmetric U-shaped architecture has become a standard structure in image segmentation tasks, such as Att U-Net \cite{Att_Unet}, FCRB U-Net \cite{cibm_cnn_2}, V-Net \cite{V_NET}, and HADCNet \cite{cibm_cnn_1}, etc. However, these methods are constrained by the receptive field of the convolutional kernel and can only model local information, lacking the capacity for global modeling. 
The DeepLab \cite{deeplab} introduced the concept of dilated convolutions to expand the receptive field of the convolutional kernel. The subsequent versions, DeepLabv3 \cite{deeplabv3} and DeepLabv3+ \cite{deeplabv3+}, used the dilated convolutions with varying dilation rates to extract multi-scale information. However, the convolutional neural networks (CNN)-based methods can only model local information in the images and lack the capability to model long-range dependencies between sequences.
\begin{figure*}[t]
\centerline{\includegraphics[width=0.6\textwidth]{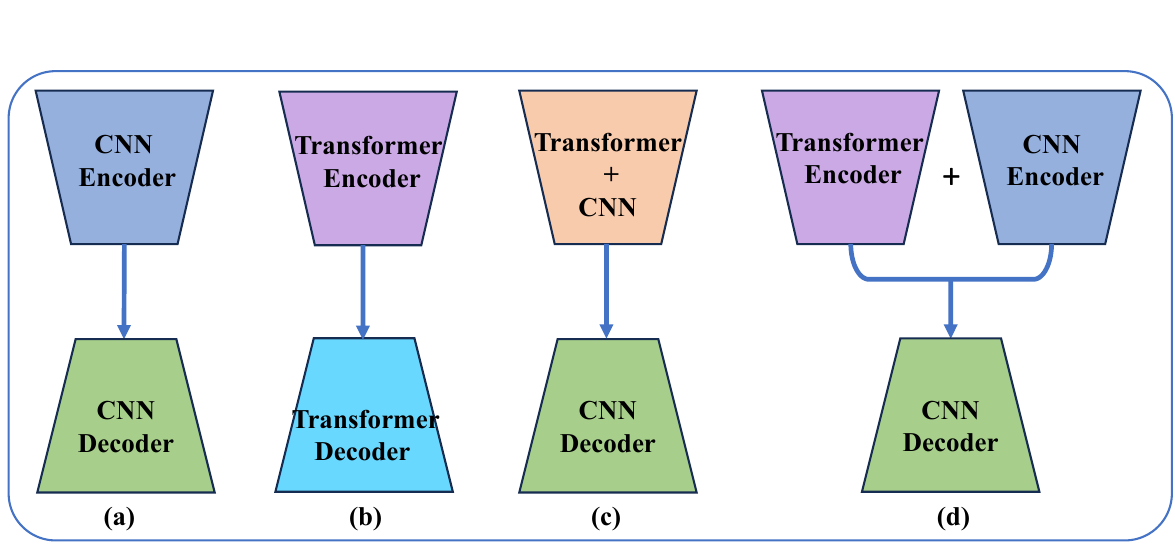}}
\caption{Conceptual comparison of the three most popular models used for medical image segmentation, where (a) classical U-Net \cite{u_net}; (b) Swin U-Net \cite{swinunet}; (c) TransUNet \cite{transunet}; (d) Our parallelized TransCNN encoder.}
\label{fig_1}
\end{figure*}

The Vision Transformer (ViT) \cite{ViT} has been introduced for computer vision tasks to address the limitations of CNN. The ViT can effectively model the relationships between sequences and sequences within images through its self-attention mechanism, enabling the extraction of global information. Various methods, such as ViTAE \cite{vitae}, Swin Transformer \cite{swin_Transformer}, and PVT \cite{PVT} have demonstrated the applicability and effectiveness of Transformers in computer vision tasks. Subsequently, the integration of Transformers into medical image segmentation has been explored. Pure Transformer architectures, such as Swin U-Net \cite{swinunet}, TransDeepLab \cite{transdeeplab}, MISSFormer \cite{MISSFormer}, and DAE-Former \cite{DAE-Former}, apply Transformer to both the encoder and decoder, allowing global features to be extracted at multiple resolutions. However, these methods overlook the advantages of CNN in learning local features, leading to inadequate extraction of detailed information from medical images. To address this issue, hybrid networks combining CNN and Transformer have been proposed for medical image segmentation, such as TransUNet \cite{transunet}, HiFormer \cite{Hiformer}, TransCeption \cite{TransCeption}, and MSRAformer \cite{MSRAformer}. These methods incorporate Transformer as the bottleneck in the encoder or as part of the decoder, often utilizing skip connections. However, despite the fusion of CNN and Transformer, these methods still do not fully utilize the feature extraction capabilities of Transformer at different resolutions. Furthermore, single encoder (see Figure \ref{fig_1}) limites the potential of CNN and Transformer for complex image processing tasks. Therefore, there is a need to further develop medical image segmentation models that effectively leverage the strengths of both Transformer and CNN, integrating their respective abilities to capture global and local information to improve medical image segmentation performance, especially for complex anatomical structures such as multi-organ and the aortic vessel tree.

Considering the distinct strengths possessed by CNN and Transformer, we propose a novel parallelized encoder architecture, called ParaTransCNN, for medical image segmentation. As shown in Figure \ref{fig_1}, there are three common configurations for the encoder and decoder: based on CNN, Transformer, or a combination of CNN and Transformer. However, approaches like TransUNet \cite{transunet} simply integrate and superimpose the features of CNN and Transformer, which cannot fully exploit the fusion effect of local and global features. Therefore, we propose a novel feature fusion technique that uses channel attention modules to fuse local and global features at different scales. Specifically, our encoder includes two branches: one utilizes a Transformer to capture global features, and the other utilizes CNN to extract local features. By using channel attention and skip connections, features at different scales are effectively fused and passed to the decoder. 
To sum up, our contributions can be concluded as follows
\begin{itemize}
\item We propose the ParaTransCNN network, a U-shaped medical image segmentation architecture involving a parallelized encoder consisting of the CNN and Transformer. Our encoder can extract local and global features of different dimensions and efficiently integrate them to provide the decoder with rich pixel-level semantic information.
\item The Transformer branch introduces a pyramid structure to learn global information at different scales, while the CNN branch uses the same downsampling strategy to learn local information. To achieve effective multi-scale feature fusion, we design a channel attention module to activate useful channel features and suppress unnecessary features.
\item We provide numerical experiments on aortic vessel tree, cardiac and multi-organ segmentation by comparing it with state-of-the-art segmentation methods. Our model shows satisfactory segmentation accuracy, especially on tiny vessel branches and small organs like the pancreas and kidneys, stomach.
\end{itemize}
\section{Related Works}
\subsection{CNN for Medical Image Segmentation}
The CNN-based methods are widely used and regarded as one of the most prominent approaches for medical image segmentation. Ronneberger et al. \cite{u_net} initially proposed the U-Net model, which consisted of an encoder and a decoder, to effectively capture local and detail features in the image. Subsequently, Oktay et al. \cite{Att_Unet} proposed the Att U-net, which improved the segmentation results by incorporating attention gates to enrich the semantic information of different dimensional feature maps. Shu et al. \cite{cibm_cnn_2} proposed a novel fully connected residual block-based FCRB U-Net for fetal cerebellum ultrasound image segmentation. It replaced the double convolution operation in the original model with fully connected residual blocks and embedded an effective channel attention module to enhance the extraction of meaningful features. Additionally, a feature reuse module was used in the decoding stage to form a fully connected decoder, making full use of deep features. Liu et al. \cite{liu2022effective} developed  a U-shape nonlocal deformable convolutional network to accurately predict the local geometry of boundaries. Besides that, the CNN-based methods have been applied to various medical image segmentation tasks, including retinal image segmentation \cite{Retinal_Images}, and skin segmentation \cite{cibm_skin_1, cibm_skin_2}, etc. These methods demonstrate promising performance in various medical image segmentation tasks and are recognized for their usability and effectiveness in implementation and training.

\subsection{Vision Transformer for Medical Image Segmentation}
In recent years, there has been extensive attention towards Transformer-based methods for medical image segmentation. The transformer leverage their self-attention mechanism to model long-range dependencies and have achieved significant breakthroughs in natural language processing. Dosovitskiy et al. \cite{ViT} proposed the vision transformer (ViT), which was the first successful application of Transformer in visual tasks, demonstrating the potential of Transformer in computer vision. Subsequently, Liu et al. \cite{swin_Transformer} proposed the window-based transformer Swin Transformer for image classification and object detection tasks. The window-based attention mechanism divides the image into small patches and performs self-attention operations on each patch, reducing the computational complexity. Wang et al. \cite{PVT} designed a vision transformer of the pyramid structure (PVT) for image classification and segmentation, where varying-sized self-attention windows were applied at distinct hierarchical levels. It facilitates the simultaneous capture of global and local contextual information, thereby facilitating enhanced comprehension of intricate image details and overarching structural components.

 Cao et al. \cite{swinunet} was the first to propose Swin U-Net, a pure transformer-based medical image segmentation model, for multi-organ segmentation tasks. It employed Swin Transformer within a U-shaped segmentation network. Azad et al. \cite{transdeeplab} proposed TransDeepLab for skin lesion segmentation, which applied different window strategies and builds upon DeepLab. Huang et al. \cite{MISSFormer} proposed MISSFormer, which aimed to utilize global information at different scales for the cardiac segmentation task, and Azad et al. \cite{TransCeption} proposed TransCeption redesigned the patch merging module in the encoder, allowing it to capture multi-scale representations within a single stage. As seen, Transformer-based methods have showcased tremendous potential in medical image segmentation tasks. 
\subsection{Combine CNN and Transformer for Medical Image Segmentation}
For medical image segmentation, Chen et al. \cite{transunet} proposed TransUNet, which was the poineer attempt to apply transformer to the encoder part of multi-organ segmentation networks. By combining local information from CNN and the encoder-decoder architecture of U-Net, TransUNet achieved significant results. Li et al. \cite{attransunet} proposed ATTransUNet, which was an improved version of TransUNet for ultrasound and histopathology image segmentation. Heidari et al. \cite{Hiformer} proposed HiFormer utilized a combination of Swin Transformer and a CNN-based encoder to design two distinct multi-scale feature representations. Additionally, a Double-Level Fusion (DLF) module was introduced in the skip connection of the encoder-decoder structure to facilitate the effective fusion of global and local features.
\begin{figure*}[t]
\centerline{\includegraphics[width=\textwidth, height=10cm]{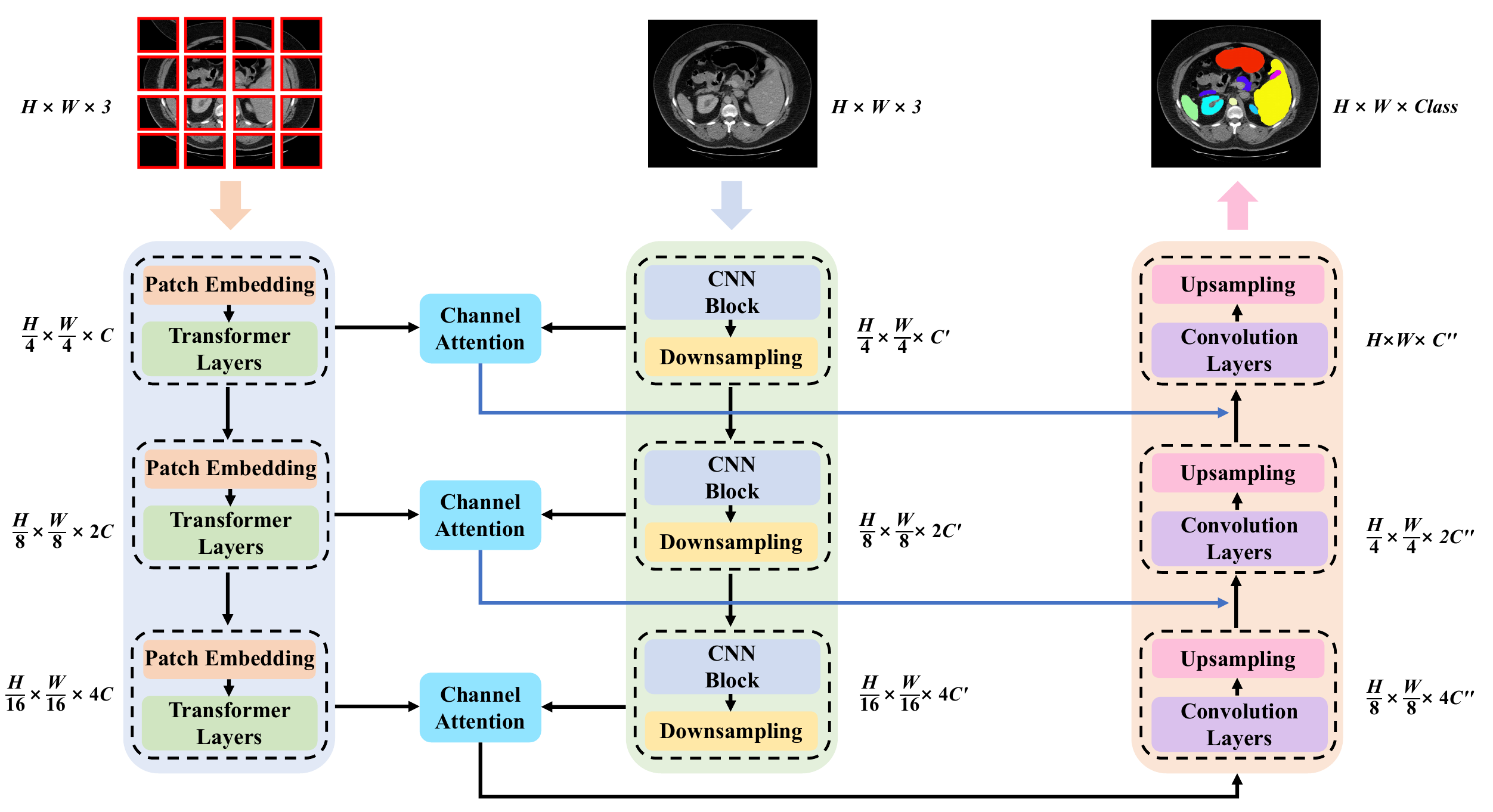}}
\caption{The overall structure of the proposed parallelized TransCNN encoder network. The parallelized encoder consists of three stages, each of which extracts local and global information at different resolutions. The Channel Attention is applied to provide rich and useful semantic information for the decoder.}
\label{ResTR}
\end{figure*}

\section{Methodology}
As shown in Figure \ref{ResTR}, we propose a parallelized encoder model for medical image segmentation. Unlike traditional segmentation models that rely solely on a single encoder branch to integrate global and local information from images, our approach incorporates two independent branches that capture semantic information from the input images separately. The overall architecture of ParaTransCNN follows the encoder-decoder paradigm similar to U-Net, consisting of an encoder, channel attention module, skip connections, and decoder. Notably, the parallelized encoder combines the advantages of ResNet \cite{resnet} and Transformer, which serve as the backbone networks. Within the Transformer component, a pyramid structure is introduced to capture global features at multiple resolutions. Additionally, a channel attention module is employed to enhance the expressive power of the parallelized encoder, enriching the extracted features and providing comprehensive guidance for the subsequent decoding process. We also utilize skip connections and decoder modules to estimate the final segmentation mask. The detailed method will be explained below.

\subsection{Parallelized TransCNN Encoder}
Inspired by the segmentation models based on CNN or Transformer mentioned earlier, we propose a parallelized encoder to address the limitations of a single-branch encoder. We introduce a pyramid structure in Transformer to obtain global feature maps at different scales. The proposed parallelized encoder includes three stages, which can extract more comprehensive and diverse feature representations.

More specifically, we employ a patch embedding layer with a patch size of 4 to operate on the 2D input image of size ${H \times W \times 3}$ (3 is channels), ensuring no overlapping patches. The resulting feature map of resolution size ${\frac{H}{4} \times\frac{W}{4} \times{C}}$ is then processed through the Transformer layer to capture its global information. Moving on to the second stage, in order to preserve detailed information within the patch embedding layer, we reduce the patch size to 2. This yields a feature map of resolution size ${\frac{H}{8} \times\frac{W}{8} \times{2C}}$, which is subsequently fed through the Transformer layer for further processing. Within the third stage, the patch size remains at 2. Using a pyramid-structured Transformer encoder, we sequentially downsample the feature maps by a factor of 4, 8, and 16. This downsampling process is essential for generating feature maps at varying scales, allowing for a broader receptive field and capturing hierarchical representations of the input image. The branch encoder can be described as follows:
\begin{equation}
    F^1_{ViT} = {TL_1(Patch\_Embed_1(X))\in\mathbb{R}^{\frac{H}{4} \times\frac{W}{4} \times{C}}}
\label{eq:eq1}
\end{equation} 
\begin{equation}
    F^2_{ViT} = {TL_2(Patch\_Embed_2(F^1_{ViT}))\in\mathbb{R}^{\frac{H}{8} \times\frac{W}{8} \times{2C}}}
\label{eq:eq2}
\end{equation}
\begin{equation}
    F^3_{ViT} = {TL_3(Patch\_Embed_3(F^2_{ViT}))\in\mathbb{R}^{\frac{H}{16} \times\frac{W}{16} \times{4C}}}
\label{eq:eq3}
\end{equation} 
respectively. Among them, $X\in\mathbb{R}^{H \times W \times\textit{3}}$ is the input image, ${Patch\_Embed_j}(\cdot)$ is the patch embedding layer of the $j$th stage and  ${TL_j}(\cdot)$ represents the features of ViT at the ${j}$th stage for ${j}\in\left\{1,2,3\right\}$.

For the other branch, we employ ResNet \cite{resnet} as the backbone network to capture the local details of the image. We perform downsampling by a factor of 4, 8, and 16, respectively, to ensure that the local information extracted by ResNet being aligned with the feature maps obtained by the transformer branch. We can describe the CNN branch as follows:
\begin{equation}
F^1_{CNN} = RL_1(X)\in\mathbb{R}^{\frac{H}{4} \times\frac{W}{4} \times{C^{\prime}}}
\label{eq:eq4}
\end{equation}
\begin{equation}
F^2_{CNN} = RL_2(F^1_{CNN})\in\mathbb{R}^{\frac{H}{8} \times\frac{W}{8} \times{2C^{\prime}}}
\label{eq:eq5}
\end{equation}
\begin{equation}
F^3_{CNN} = RL_3(F^2_{CNN})\in\mathbb{R}^{\frac{H}{16} \times\frac{W}{16} \times{4C^{\prime}}}
\label{eq:eq6}
\end{equation}
respectively. Among them, $X\in\mathbb{R}^{H \times W \times\textit{3}}$ is the input image, ${RL_i}(\cdot)$ represents the features of CNN branch at the ${i}$th stage for ${i}\in\left\{1,2,3\right\}$.

\subsection{Channel Attention Module}
\begin{figure*}[t]
\centerline{\includegraphics[width=\textwidth]{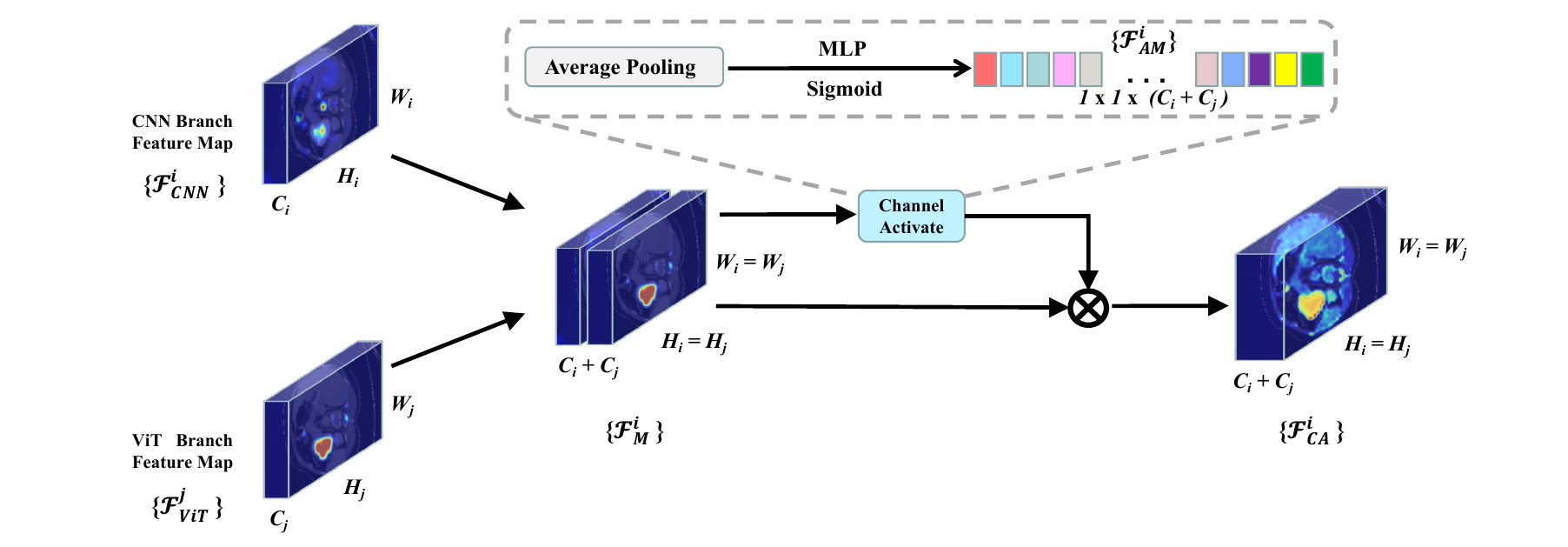}}
\caption{The channel attention module. By taking organ segmentation as an example, it can be seen that channel weighting of the features of the parallelized encoder can enrich more semantic information.}
\label{channel}
\end{figure*}

We introduce a channel attention module (see Figure \ref{channel}) to combine the local and global information obtained from the CNN and Transformer branches, so that valuable information can be passed from the parallelized encoder to the decoder, resulting in pixel-level segmentation results. As simple convolution operations are inadequate for effectively fusing local and global features, we weigh the feature channels according to their information content. By activating channels that contribute to segmentation results and suppressing irrelevant ones, we achieve efficient feature fusion utilizing local and global information from the encoder. More specifically, we concatenate the features ${F^i_{CNN}}\in\mathbb{R}^{{H_i} \times {W_i} \times{C_i}}$ from the CNN branch and ${F^j_{ViT}}\in\mathbb{R}^{{H_j} \times {W_j} \times{C_j}}$ from the transformer branch to obtain the merged feature ${F^i_{M}}\in\mathbb{R}^{{H_i} \times {W_i} \times{(C_i + C_j)}}$, which is given as:
\begin{equation}
F^i_{M} = Concat[F^i_{CNN}, F^j_{ViT}]
\label{eq:eq7}
\end{equation}
Next, we utilize the average pooling to acquire a representation of the channel information and then apply a multi-layer perceptron and a sigmoid activation function to perform a non-linear transformation on the channel representation, yielding a channel attention map ${F^i_{AM}}\in\mathbb{R}^{\textit{1} \times \textit{1} \times{(C_i + C_j)}}$. 
Subsequently, we perform an element-wise multiplication of the channel attention map ${F^i_{AM}}$ with the merged feature ${F^i_{M}}$ assigning varying weights to different channels in order to activate useful information from the parallelized encoder and suppress irrelevant information. This results in ${F^i_{CA}} \in\mathbb{R}^{{H_i} \times {W_i} \times{(C_i + C_j)}}$, as follows:
\begin{equation}
F^i_{CA} = ({F^i_{AM}} 
{\scriptstyle{\bigotimes}} {F^i_{M}})
\label{eq:eq8}
\end{equation}
As illustrated in Figure \ref{channel}, the features that undergo processing by the channel attention module hold more rich semantic information than those from any single branch.

\subsection{Decoder}
By implementing a parallelized encoder and channel attention module, we achieve multi-scale feature representations with enriched semantic information. Similar to TransUNet, we use skip connections to link the low-resolution features to high-resolution features, which are then passed to the decoder to generate the final segmentation mask. The decoder employs convolutional layers to extract the combined multi-scale features, including $3\times3$ convolutions, Batch Normalization \cite{batch_nor}, and ReLU layers. The upsampling operations are performed using transpose convolutional layers, sequentially attaining 2$\times$, 2$\times$, and 4$\times$ upsampling. Notably, our decoder does not incorporate any extra techniques or strategies such as the residual connections \cite{Unetr} or deep supervision \cite{Polyp_PVT}. Furthermore, we observe that incorporating a transformer as the decoder does not yield any significant performance improvement.

\subsection{Loss Function}
We use both the Dice loss and Cross Entropy as the loss function for training our ParaTransCNN network, which is given as follows:
\begin{equation}
{\mathcal{L}} = {\lambda_{1}}{\mathcal{L}_{Dice}} + {\lambda_{2}}{\mathcal{L}_{CE}}
\label{eq:eq9}
\end{equation}
where ${\mathcal{L}}$ is the total loss, ${\lambda_{1}}$ and ${\lambda_{2}}$ represent different weight coefficients, being experimentally set as 0.5 and 0.5, respectively.

\section{Model Implementation}
\subsection{Datasets}

\subsubsection{Aortic Vessel Tree (AVT) Segmentation}
We use the 56 CTA scans from SEG.A. challenge 2023. This SEG.A. dataset consists of three centers, including the KiTS Grand Challenge (K dataset with 20 cases), the Rider Lung CT dataset (R dataset with 18 cases), and the Dongyang Hospital (D dataset with 18 cases) \cite{AVT}. We randomly selected 38 cases as training sets (15044 slices), and 18 cases to test. Besides, we resample all slices to 1mm $\times$ 1mm, with the HU value of [0,1].

\subsubsection{Automated Cardiac Diagnosis Challenge (ACDC) Segmentation}
The ACDC challenge collects exams from different patients acquired from MRI scanners. The cine MR images were acquired in breath hold, and a series of short-axis slices covered the heart from the base to the apex of the left ventricle, with a slice thickness of 5 to 8 mm. The short-axis in-plane spatial resolution goes from 0.83 to 1.75 mm2/pixel \cite{ACDC}. 
Each patient scan is manually annotated with ground truth for left ventricle (LV), right ventricle (RV), and myocardium (MYO). We use 100 training cases, and 50 cases for testing.

\subsubsection{Synapse Multi-Organ Segmentation} The multi-organ Synapse dataset consists of 30 abdominal CT scans (12 cases to test) with 3779 axial contrast-enhanced clinical CT images; see \cite{Multi-Organ} . Each volumetric sample is composed of 85 $\thicksim$ 198 slices of the same size of 512 $\times$ 512 pixels in the entire 3D data. All slices were resampled to 1mm x 1mm, with the HU value of [0,1].
\begin{table*}[H]
    \centering
    \renewcommand{\arraystretch}{1.3}
    \caption{Ablation study on the token dim of ViT branch on the \textit{Synapse} dataset.}
        \label{a}
        \resizebox{\textwidth}{!}{
    	\begin{tabular}{lcccc*{8}c}
           \hline Model & Dim (C) & Layer & DSC(\%)$\uparrow$ & HD$\downarrow$ & Aorta & Gallbladder & Kidney(L) & Kidney(R) & Liver & Pancreas & Spleen & Stomach \\
    		\hline 
    		ParaTransCNN-Small & 64 & [3,3,3] & 82.93 & 16.44 & \textbf{89.39} & \textbf{69.24} & 86.74 & 81.63 & 94.88 & 67.87 & 91.58 & 82.13\\
    		ParaTransCNN-Base & 192 & [3,3,3] & 82.01 &20.45 & 88.58 & 66.86 & 85.20 & 78.49 & \textbf{95.37} & 67.33 & 91.56 & 82.69 \\
    		\textbf{ParaTransCNN-Medium} & \textbf{320} & \textbf{[3,3,3]} & \textbf{83.86} & \textbf{15.86} & 88.12 & 68.97 & \textbf{87.99} & \textbf{83.84} & 95.01 & \textbf{69.79} & \textbf{92.71} & \textbf{84.43}\\
    		ParaTransCNN-Large & 512 & [3,3,3] & 81.87 & 17.30 & 89.10 & 68.18 & 86.13 & 80.33 & 94.48 & 67.57 & 89.84 & 79.31  \\
    		\hline
      \end{tabular}}
\end{table*}    
\begin{table*}[H]
      \centering
    \renewcommand{\arraystretch}{1.3}
        \caption{Ablation study on the Transformer layer numbers of ViT branch on the \textit{Synapse} dataset.}
        \label{b}
        \resizebox{\textwidth}{!}{
        \begin{tabular}{lcccc*{8}c}
            \hline Model & C & Layer & DSC(\%)~$\uparrow$& HD~$\downarrow$ & Aorta & Gallbladder & Kidney(L) & Kidney(R) & Liver & Pancreas & Spleen & Stomach \\
    		\hline 
    		ParaTransCNN-Medium (Level 1) & 320 & [2,3,3] & 81.68 & 17.16 & 88.68 & 67.07 & 86.52 & 81.36 & 95.07 & 65.06 & 90.58 & 79.13 \\
    		\textbf{ParaTransCNN-Medium (Level 2)} & \textbf{320} & \textbf{[3,3,3]} & \textbf{83.86} & 15.86 & 88.12 & \textbf{68.97} & \textbf{87.99} & \textbf{83.84} & 95.01 & \textbf{69.79} & \textbf{92.71} & \textbf{84.43}\\
    		ParaTransCNN-Medium (Level 3) & 320 & [3,3,4] & 82.41 & \textbf{13.00} & 88.55 & 67.55 & 86.63 & 81.31 & 94.88 & 67.36 & 91.19 & 81.85\\
    		ParaTransCNN-Medium (Level 4) & 320 & [3,6,3] & 80.80 & 26.09 & \textbf{88.79} & 66.61 &82.18 & 74.68 & \textbf{95.11} & 64.27 & 91.82 & 82.96 \\
    		\hline
        \end{tabular}
        }
\end{table*}

\subsection{Implementation Details}
All ParaTransCNN experiments are trained on a single NVIDIA RTX 3090 GPU. We use the SGD optimizer to optimize the network parameters, where the batch size is set at 4 and the total number of 150 epochs is used. The initial learning rate is set as 0.01 and the learning rate decreases with the use of the power of 0.9 for training. All data have the unified input size of 224 $\times$ 224. We apply simple data augmentations including random rotation and flip. Both Dice Similarity Coefficient (DSC) and 95$\%$ Hausdorff Distance (HD) are used as the rating indicators for segmentation accuracy.

\section{Experimental Results}
\subsection{Ablation Study}
\subsubsection{Parameter Ablation}
Since the Transformer model assesses the relationships between sequences, it is essential to choose an appropriate sequence length for patch representation to effectively capture global information. To identify the optimal token dimension (denoted by C), we create a variety of ParaTransCNN models with different scales: ParaTransCNN-Small, ParaTransCNN-Base, ParaTransCNN-Medium, and ParaTransCNN-Large. These models correspond to token dimensions of 64, 192, 320, and 512, respectively. To ensure consistency, all three stages of the Transformer Layer were set to 3. The results, displayed in Table \ref{a}, reveal that the model attains the best segmentation performance with a token dimension of 320. Compared to ParaTransCNN-Large, ParaTransCNN-Medium exhibits a 1.99\% higher Dice Similarity Coefficient (DSC) and also records the lowest 95$\%$ Hausdorff distance (HD).

\begin{figure*}[H]
\centerline{\includegraphics[width=\textwidth]{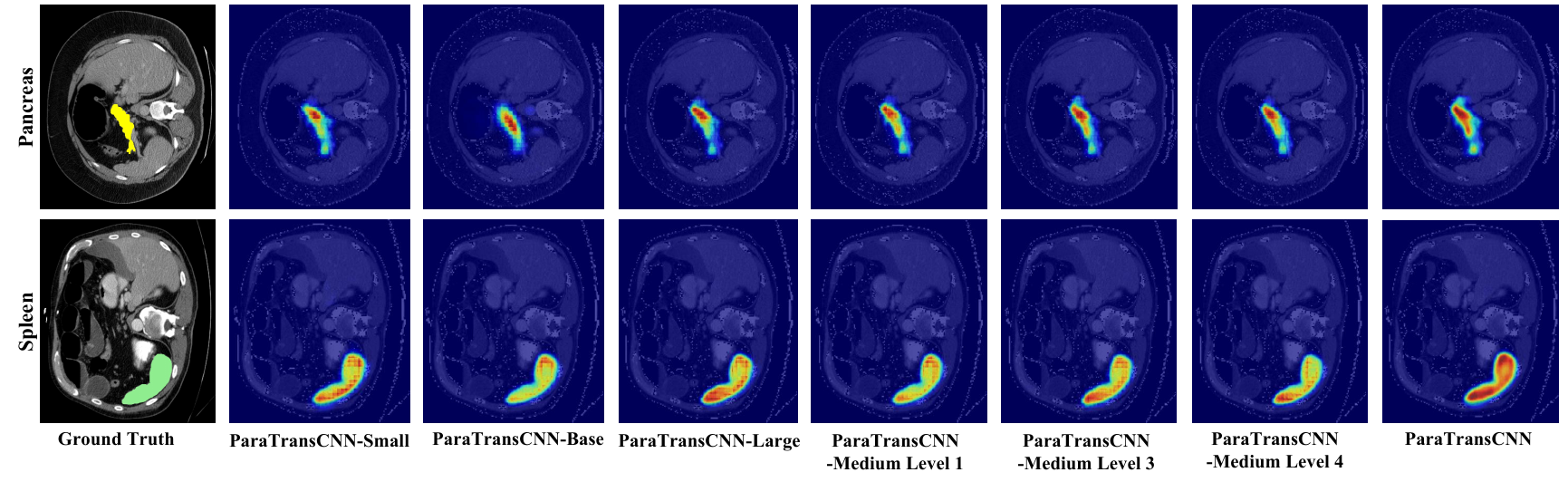}}
\caption{Visual comparison of attention map from different token dim and transformer layer numbers in the ViT branch build models in small organs of the pancreas, and spleen. The brighter the color, the higher the attention paid to the features of the area. ParaTransCNN-Medium is the same as ParaTransCNN-Medium (Level 2), it is our ParaTransCNN.}
\label{xiaorong_1}
\end{figure*}

\begin{figure*}
\centering
\begin{minipage}[t]{0.48\textwidth}
\centering
\includegraphics[width=\textwidth]{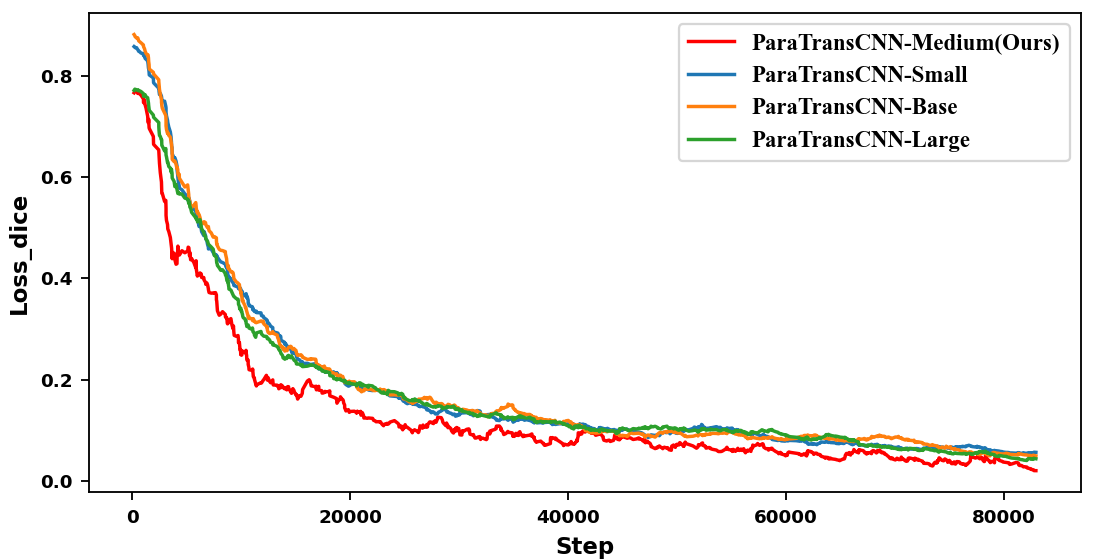}
\end{minipage}
\begin{minipage}[t]{0.48\textwidth}
\includegraphics[width=\textwidth]{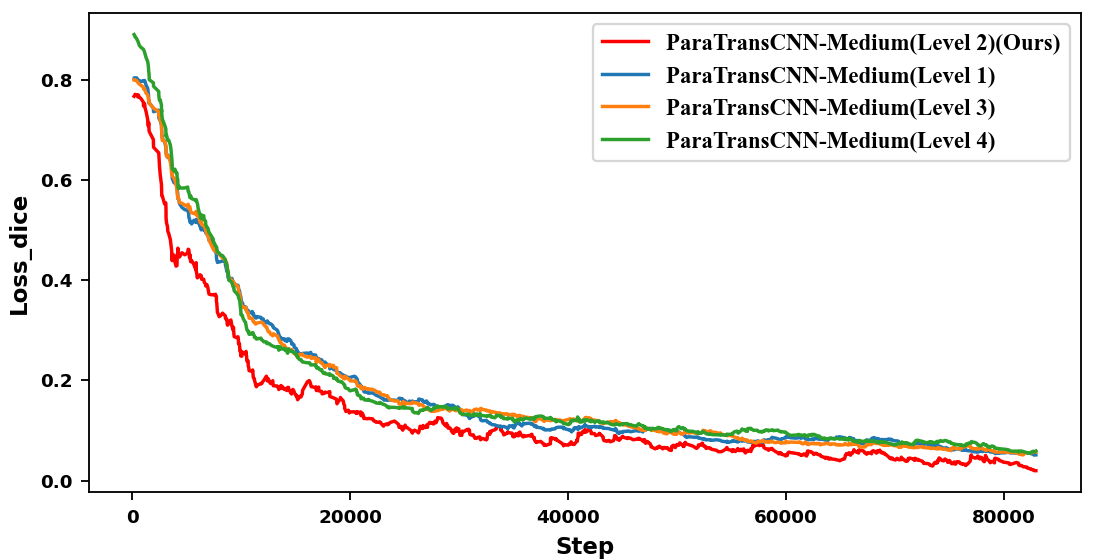}
\end{minipage}
\caption{Loss curves under ParaTransCNN models of different scales and ParaTransCNN-Medium of different levels.}
\label{loss_log}
\end{figure*}

Both the token dimension and the number of layers in the Transformer architecture are crucial for capturing global information. The token dimension influences the ability to encode and process long-range dependencies, while the Transformer layers enhance the model's capacity to represent complex relationships among tokens. To strike a balance between computational efficiency and performance, it is essential to determine the optimal number of stacked layers. We examine the impact of Transformer layers on the ParaTransCNN-Medium model by conducting experiments with varying levels from level 1 to level 4. The results, displayed in Table \ref{b}, show that using three layers per stage in the Transformer architecture achieves the best segmentation performance. Notably, the level 2 configuration improves the DSC by 3.06\% compared to the level 4. Figure \ref{xiaorong_1} presents the feature attention maps of the model under different parameter settings, illustrating that our approach emphasizes smaller organs like the pancreas and spleen more effectively. These experimental results validate the rationale behind our token dimension and Transformer layer settings in the ParaTransCNN-Medium model. In addition, the convergence analysis in Figure \ref{loss_log} show that our parameter settings achieve lower loss values after the same number of iterations under the same training conditions.
\begin{table*}[H]
    \centering
    \renewcommand{\arraystretch}{1.3}
    \caption{The ablation study on the model structure of our ParaTransCNN on the \textit{Synapse} dataset, where 'Patch Overlap' represents the overlapping between patches in each stage of the patch embedding layer in the Transformer branch, '4 Downsampling Stages' represents a down-sampling rate of 32 is used in the fourth stage of the parallelized encoder, 'W/O Pyramid of Transformer' and 'W/O Channel Attention' represent a Transformer branch without pyramid structure and without channel attention module,  respectively.}
    \label{jiegou}
    \resizebox{\textwidth}{!}{
    	\begin{tabular}{lcc*{8}c}
    		\hline Model & DSC(\%)~$\uparrow$ & HD~$\downarrow$  & Aorta & Gallbladder & Kidney(L) & Kidney(R) & Liver & Pancreas & Spleen & Stomach  \\
    		\hline 
    		Patch Overlap & 82.77 & 19.86 & 88.68 & \textbf{72.89} & 85.00 & 78.84 & \textbf{95.11} & 67.80 & 91.44 & 82.40\\
    		4 Downsampling Stages & 80.59 & 16.30 & 86.70 & 67.26 & 83.54 & 78.77 & 95.06 & 66.70 & 90.15 & 76.57\\
    		W/O Pyramid of Transformer & 77.15 & 33.11 & 86.33& 59.77 &80.71 &72.86 & 94.15 & 56.08 & 88.50 & 78.84 \\
    		W/O Channel Attention & 81.58 & 25.68 & \textbf{89.06} & 71.76 & 85.53 & 76.00 & 94.75 & 65.78 & 90.76 & 79.03  \\
    		\textbf{ParaTransCNN} & \textbf{83.86} & \textbf{15.86} & 88.12 & 68.97 & \textbf{87.99} & \textbf{83.84} & 95.01 & \textbf{69.79} & \textbf{92.71} & \textbf{84.43} \\
    		\hline
    	\end{tabular}
    }
\end{table*}
\begin{figure*}[H]
\centerline{\includegraphics[width=1.1\textwidth]{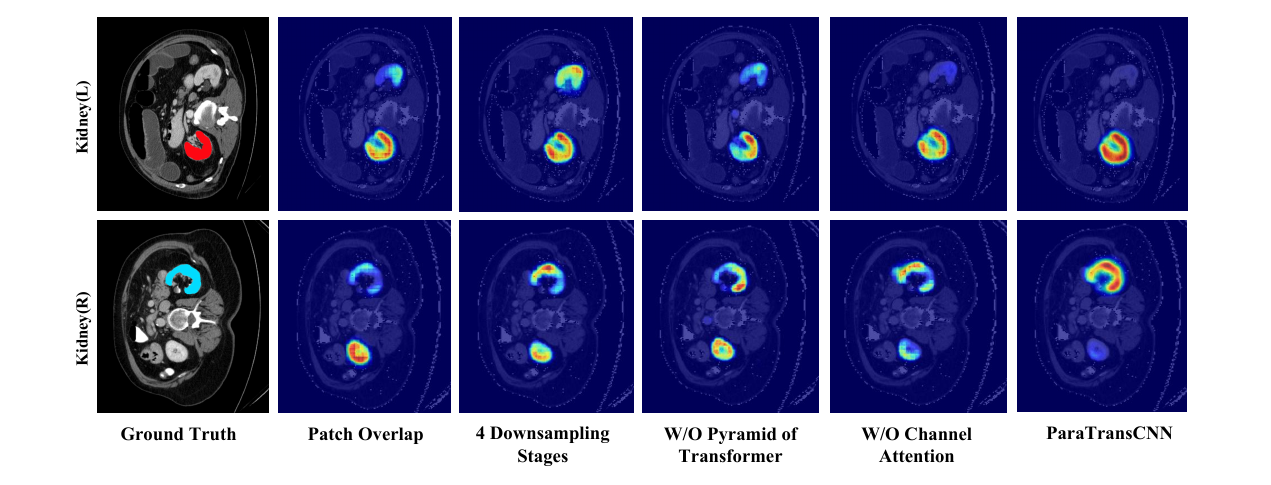}}
\caption{Visual comparison of the attention map obtained by different models. It can be found that in the visualization of the left and right kidneys, the other four structures of interest are wrong, and the attention of the correct area of attention is not significant.}
\label{xiaorong_2}
\end{figure*}

\begin{table*}[t]
\centering
 \renewcommand{\arraystretch}{1.3}
	\caption{Comparison results with the state-of-the-art methods on the \textit{AVT} dataset.} \label{sota_avt}
 \resizebox{\textwidth}{!}{
	\begin{tabular}{lccccccccc}
		\hline
		 Methods & \multicolumn{2}{c}{D dataset} & \multicolumn{2}{c}{R dataset} & \multicolumn{2}{c}{K dataset} & \multicolumn{2}{c}{All dataset}\\ 
       \cline{2-9}
		 & DSC(\%)~$\uparrow$ & HD~$\downarrow$ & DSC(\%)~$\uparrow$ & HD~$\downarrow$ & DSC(\%)~$\uparrow$ & HD~$\downarrow$&DSC(\%)~$\uparrow$ & HD~$\downarrow$ \\ 
         \hline
		 U-Net \cite{u_net} & 93.13 & 2.65 & 76.32 & 10.34 & 79.22 & 8.83 & 82.89 & 7.27\\
           Att-UNet \cite{Att_Unet} & 92.69 & 3.58 & 70.15 & 15.18 & 77.91 & 9.58 & 80.25 & 9.45\\
             TransUNet \cite{transunet} & 93.59 & 1.21 & 74.73 & 23.47 & 80.67 & 9.61 & 82.99 & 11.43\\
             Swin U-Net \cite{swinunet} & 92.02 & 2.19 & 69.41 & 26.48 & 77.45 & 12.62 & 79.62& 13.76\\
             TransDeepLab \cite{transdeeplab} & 92.97 & 1.90 & 71.06 & 24.88 & 75.15 & 12.06 & 79.72 & 12.95\\
             HiFormer \cite{Hiformer} & 93.35 & 1.41 & 75.56 & 11.19 & 78.69 & 9.45 & 82.53 & 7.35\\
             MISSFormer \cite{MISSFormer} & 92.28 & 2.67 & 69.20 & 21.62 & 75.81 & 12.45 & 79.09 & 12.25\\
             TransCeption \cite{TransCeption} & 93.70 & 1.21 & 75.54 & 17.92 & 79.37 & 11.97 & 82.87 & 10.33\\
             DAE-Former \cite{DAE-Former} & \textbf{93.96} & 1.21 & 78.85 & 14.32 & 82.25 & 11.37 & 85.02 & 8.96\\
             \textbf{ParaTransCNN(Ours)} & 93.91 & \textbf{1.07} & \textbf{83.01} & \textbf{7.75} & \textbf{86.55} & \textbf{5.27} & \textbf{87.82} &\textbf{4.70}\\
             \hline
	\end{tabular}}
\end{table*}
\begin{figure*}[t]
\centerline{\includegraphics[width=\textwidth]{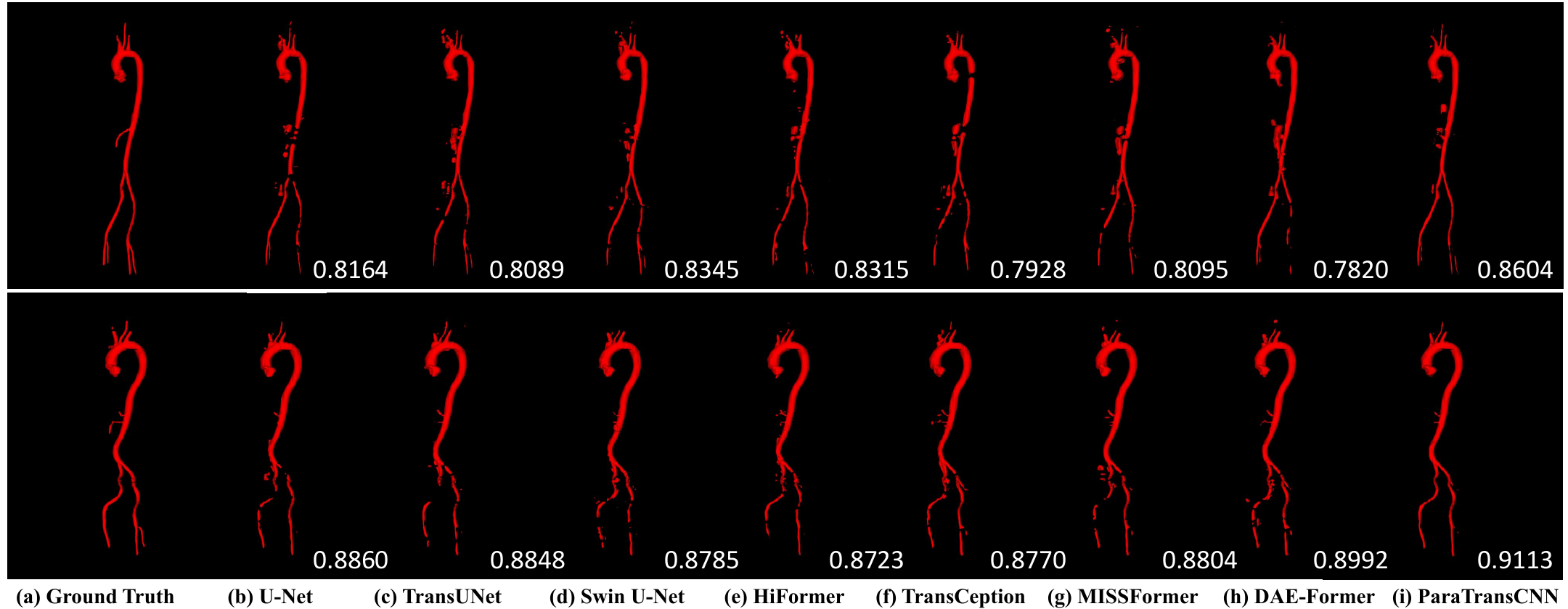}}
\caption{Qualitative results of different models on the AVT dataset. It can be seen that our ParaTransCNN has the best predictive effect continuity on aortic vessel trees and the fewest faults on small vessel branches. The value in the lower right corner represents the DSC value for the predicted volume.}
\label{sota_Avt}
\end{figure*}

\subsubsection{Structure Ablation}
To demonstrate the structural effectiveness of ParaTransCNN, we perform a series of structural ablation studies. First, we explore various patch overlap strategies in the Patch Embedding layer of each stage, and Table \ref{jiegou} shows that patch overlap does not offer any significant benefits for ParaTransCNN. Next, in an attempt to enhance the model's learning capacity, we increase the depth of the parallelized encoder by using a downsampling rate of 32 in the fourth stage. However, the DSC metric in Table \ref{jiegou} indicates a performance of only 80.59\%, which is 3.27\% lower than that of ParaTransCNN. Additionally, we examine the segmentation performance of the models without pyramid structures in the Transformer branch and without the channel attention module. Table \ref{jiegou} reveals that ParaTransCNN does not need patch overlap or an exceptionally deep encoder. Instead, incorporating a pyramid structure into the Transformer branch and activating the features of each stage in the parallelized encoder leads to the best segmentation effect. Figure \ref{xiaorong_2} displays the attention effect of different model structures on the kidneys (Left, Right). It can be observed that other methods have a lower focus on the left kidney and right kidney, and there are also instances of organ attention confusion, such as in columns 2 to 4. In contrast, our method not only avoids incorrect judgments but also exhibits higher attention performance on the left kidney and right kidney compared to other methods.

\subsection{Comparison with State-of-the-Art Methods}
\subsubsection{Results on Aortic Vessel Tree Segmentation}
In Table \ref{sota_avt}, we compare our ParaTransCNN with the state-of-the-art models on the aortic vessel tree dataset. To assess the segmentation performance of various methods on data from three different centers, we randomly selected 6 cases each from the D, R, and K datasets as test datasets, totaling 18 cases. The results indicate that the ParaTransCNN approach achieves the best performance in terms of Dice Similarity Coefficient (DSC) and 95$\%$ Hausdorff Distance (HD) across all test cases, with values of 87.82\% and 4.70, respectively, outperforming the second-best method by 2.8\% and 2.57. Moreover, our ParaTransCNN method outperforms the second-best approach by 4.16\% and 4.3\% in terms of DSC for the R and K datasets, respectively. The visualizations presented in Figure \ref{sota_Avt} clearly demonstrate that our method produces superior details and continuous structures in the segmentation of the aortic vessel tree. In contrast, other methods exhibit discontinuities and missed detections on small vessel branches, such as U-Net, which is truncated at the junction of blood vessels.

The predicted masks by TransUNet and Swin U-Net lack continuity on longer branches of the vessel, while TransCeption and MISSFormer exhibit not only discontinuities in the sub-branches of the vessels but also longer gaps in the aorta. These results demonstrate the excellent segmentation accuracy of our ParaTransCNN approach and its ability to generalize across multi-center data for aortic vessel tree segmentation.
\begin{table*}[t]
    \centering
    \renewcommand{\arraystretch}{1.3}
    \caption{Comparison with the state-of-the-art methods on the \textit{ACDC} dataset.}
    \label{ACDC}
    \resizebox{0.95\textwidth}{!}{
    	\begin{tabular}{lcc*{5}c}
                \hline
     Methods &DSC(\%)~$\uparrow$ & HD~$\downarrow$& RV & Myo & LV & Params(M) &  FLOPs(G) \\
    		\hline 
                U-Net \cite{u_net} & 90.68 & 1.51 & 91.65 & 84.82 & 95.57 & 34.53 & 50.19  \\
    		  Att-UNet \cite{Att_Unet} & 90.90 & 1.52 & 91.33 & 85.79 & \textbf{95.58} & 34.88 & 51.04\\
    		  TransUNet \cite{transunet} & 90.79 & \textbf{1.08} &91.56 & 85.24 & 95.18 & 105.28 & 24.68 \\
    		  Swin U-Net \cite{swinunet} & 90.25 & 1.32 & 90.67 & 84.92 & 95.16 & 27.17 & \textbf{5.92} \\
       TransDeepLab \cite{transdeeplab} & 90.41 & 1.12 & 91.45 & 84.54 &95.23 & \textbf{21.14} & 15.68\\
             HiFormer \cite{Hiformer} &  90.12 & 2.15 & 91.06 & 84.54 & 94.77 & 25.51 & 17.76 \\
    		 MISSFormer \cite{MISSFormer} &86.78 & 2.32 & 86.04& 80.44 & 93.84 & 42.46 & 7.25 \\
    		 TransCeption \cite{TransCeption} & 88.47 & 1.78 & 87.88 & 82.87 & 94.66 & 47.32 & 14.96\\
    	DAE-Former \cite{DAE-Former} & 89.78 & 1.91 & 89.91 & 84.38 & 95.04 & 48.07 & 26.11\\
            \textbf{ParaTransCNN(Ours)} & \textbf{91.31} & 1.16 & \textbf{92.76} & \textbf{85.84} & 95.34 & 42.70 & 50.39\\
                \hline
    	\end{tabular}}
\end{table*}
\begin{figure*}[t]
\centerline{\includegraphics[width=\textwidth]{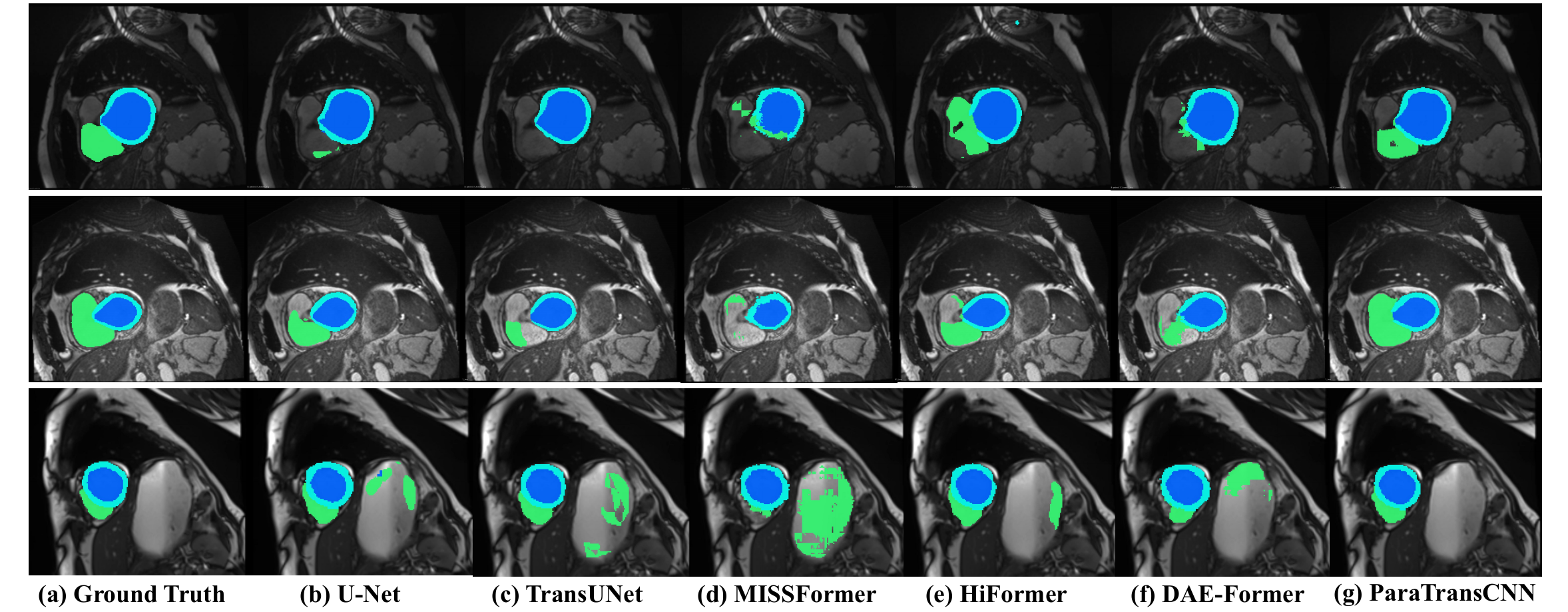}}
\caption{Qualitative results of different models on the ACDC dataset, where the green represents the right ventricle, the blue represents the left ventricle, and the light blue represents the myocardium.}
\label{sota_ACDC}
\end{figure*}
\subsubsection{Results on ACDC Segmentation}
Similarly, we train our ParaTransCNN on the ACDC dataset to perform cardiac segmentation. The experimental results are summarized in Table \ref{ACDC}. In terms of DSC, our method outperforms Swin U-Net by 1.06$\%$ and DAE-Former by 1.53$\%$. The visual results are presented in Figure \ref{sota_ACDC}. We can observe that our ParaTransCNN achieves the best segmentation performance on the cardiac segmentation, particularly for the right ventricle with low clarity. Compared to other methods, our model can yield ideal segmentation results, especially for myocardial segmentation, accurately locating the myocardial boundaries. On the other hand, other methods fail to accurately capture the topological structure of the myocardium. It demonstrates the effectiveness of the parallelized encoder in efficiently extracting local contextual and global information.
\begin{table*}[h]
    \centering
    \renewcommand{\arraystretch}{1.3}
    \caption{Comparison with the state-of-the-art methods on the \textit{Synapse} dataset.}
    \label{BTCV}
    \resizebox{\textwidth}{!}{
    	\begin{tabular}{lcc*{8}c}
                \hline
     Methods & DSC(\%)~$\uparrow$ & HD~$\downarrow$ & Aorta & Gallbladder & Kidney(L) & Kidney(R) & Liver & Pancreas & Spleen & Stomach \\
                     \hline
    		  V-Net \cite{V_NET} & 68.81 &- &75.34 &51.87 &77.10 &80.75 &87.84 &40.05 &80.56 &56.98\\
    		  DARR \cite{DARR} & 69.77 &- &74.74 &53.77 &72.31 &73.24 &94.08 &54.18 &89.90 &45.96\\
    		  R50 U-Net \cite{transunet} & 74.68 & 36.87 & 87.47 & 66.36 & 80.60 & 78.19 & 93.74 & 56.90 & 85.87 & 74.16  \\
         U-Net \cite{u_net} & 76.85 & 39.70 & 89.07 & 69.72 & 77.77 & 68.60 & 93.43 & 53.98 & 86.67 & 75.58  \\
    		  R50 Att-UNet \cite{transunet} & 75.57 & 36.97 & 55.92 & 63.91 & 79.20 & 72.71 & 93.56 & 49.37 & 87.19 & 74.95 \\
    		  Att-UNet \cite{Att_Unet} & 77.77 & 36.02 & \textbf{89.55} & 68.88 & 77.98 & 71.11 & 93.57 & 58.04 & 87.30 & 75.75 \\
    	      R50 ViT \cite{transunet} & 71.29 & 32.87 & 73.73 & 55.13 & 75.80 & 72.20 & 91.51 & 45.99 & 81.99 & 73.95 \\
    		  TransUNet \cite{transunet} & 77.48 & 31.69 & 87.23 & 63.13 & 81.87 & 77.02 & 94.08 & 55.86 & 85.08 & 75.62 \\
               TransNorm \cite{Transnorm} & 78.40 & 30.25 & 86.23 & 65.10 & 82.18 & 78.63 & 94.22 & 55.34 & 89.50 & 76.01 \\
    		  Swin U-Net \cite{swinunet} & 79.13 & 21.55 & 85.47 & 66.53 & 83.28 & 79.61 & 94.29 & 56.58 & 90.66 & 76.60\\
       TransDeepLab \cite{transdeeplab} & 80.16 & 21.25 & 86.04 & 69.16 &84.08 & 79.88 & 93.53 & 61.19 & 89.00 & 78.40\\
             HiFormer \cite{Hiformer} &  80.39 & \textbf{14.70} & 86.21 & 65.69 & 85.23 & 79.77 & 94.61 & 59.52 & 90.99 & 81.08\\
    		 MISSFormer \cite{MISSFormer} &81.96 & 18.20 & 86.99 & 68.65 & 85.21 & 82.00 & 94.41 & 65.67 & 91.92 & 80.81\\
    		 TransCeption \cite{TransCeption} & 82.24 & 20.89 & 87.60 & 71.82 & 86.23 & 80.29&\textbf{95.01} & 65.27 & 91.68 & 80.02\\
    	DAE-Former \cite{DAE-Former} & 82.43 & 17.46 & 88.96 & \textbf{72.30} & 86.08 & 80.88 & 94.98 & 65.12 & 91.94 & 79.19\\
            \textbf{ParaTransCNN(Ours)} & \textbf{83.86} & 15.86 & 88.12 & 68.97 & \textbf{87.99} & \textbf{83.84} & \textbf{95.01} & \textbf{69.79} & \textbf{92.71} & \textbf{84.43}\\
                \hline
    	\end{tabular}
}
\end{table*}
\begin{figure*}[h]
\centerline{\includegraphics[width=\textwidth]{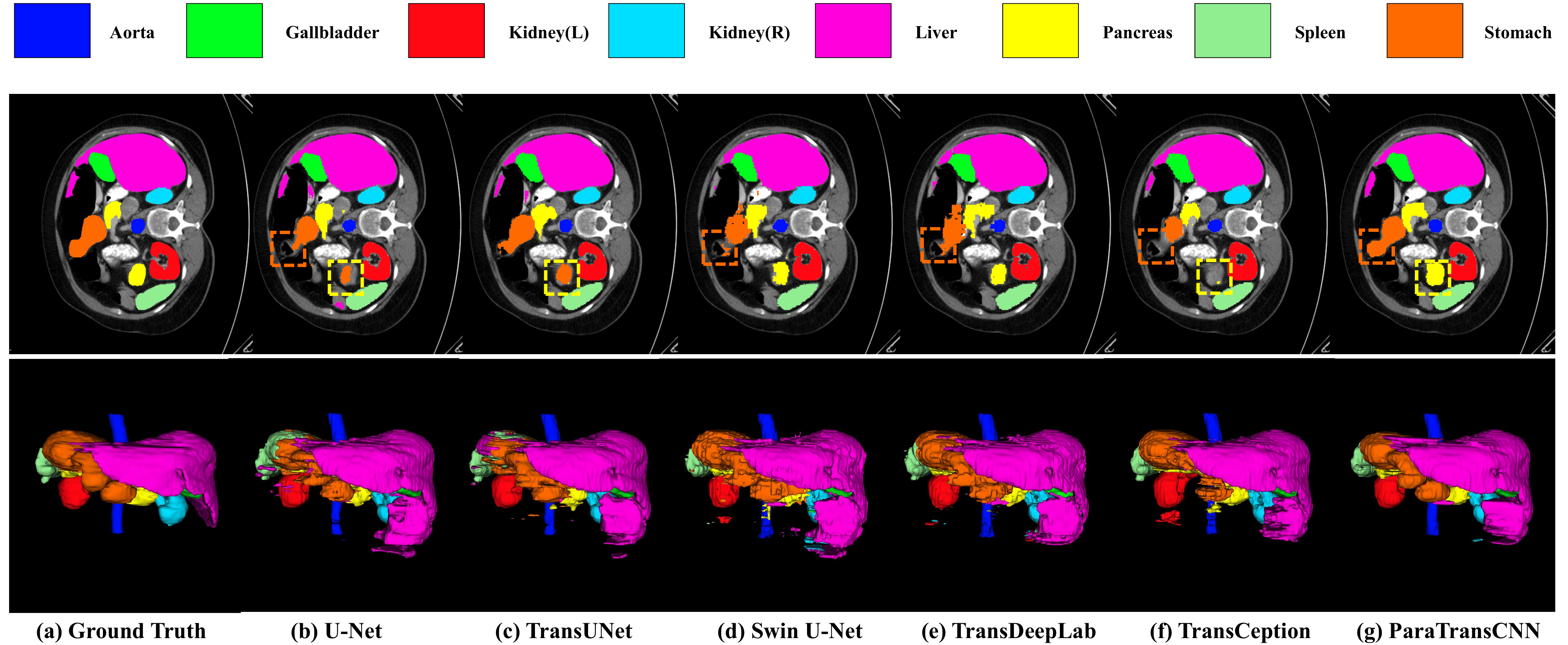}}
\caption{Qualitative results of different models on the Synapse dataset, where (a) to (g) are 2D and 3D visual representations of Ground Truth, U-Net,
TransUNet, Swin U-Net, TransDeepLab, TransCeption, and our ParaTransCNN, respectively. As can be seen from the first row, the organs in the dotted box have been missed or misdirected. In contrast, ParaTransCNN predicts best on the pancreas and stomach. In the 3D view display in the second row, the prediction of our ParaTransCNN is also the most accurate and complete.}
\label{sota_btcv}
\end{figure*}
\subsubsection{Results on Synapse Multi-Organ Segmentation}
To assess the multi-organ segmentation performance of ParaTransCNN, we compare it with CNN-based segmentation networks like R50 U-Net, U-Net \cite{u_net}, Att-UNet \cite{Att_Unet}, pure Transformer segmentation network systems such as Swin U-Net \cite{swinunet}, TransDeepLap \cite{transdeeplab}, MISSFormer \cite{MISSFormer}, DAE-Former \cite{DAE-Former}, and the mixed CNN and Transformer models like TransUNet \cite{transunet}, HiFormer \cite{Hiformer}, TransCeption \cite{TransCeption}. In Table \ref{BTCV}, our ParaTransCNN achieves the best performance in terms of the Dice Similarity Coefficient (DSC) at 83.86\% and ranks second in 95$\%$ Hausdorff Distance (HD) at 15.86. Notably, ParaTransCNN shows significant improvements in DSC for pancreas and stomach segmentation, with respective increases of 4.12\% and 3.35\% compared to other methods. It is worth noting that we do not employ transfer learning techniques such as Swin Transformer \cite{swin_Transformer} or PVT \cite{PVT} in the Transformer module, emphasizing the superiority of our parallelized encoder in feature extraction and the effectiveness of feature fusion at each stage. 

Moreover, we visualize the segmentation performance of ParaTransCNN for each organ using 2D and 3D representations in Figure \ref{sota_btcv}. As can be seen, in the task of segmenting images of small organs like the pancreas and stomach, traditional U-Net methods have limitations, which cannot accurately predict the location of the stomach region and also make errors in predicting the pancreas region. Comparison methods such as Swin U-Net and TransDeepLab also fail to accurately infer the complete stomach region and produce rough edge predictions. Both TransCeption and TransUNet showcase missed detection in the stomach and pancreas regions. On the other hand, our ParaTransCNN method can accurately identify the pancreas and stomach in image segmentation tasks. The predicted regions are not only complete but also have smooth edges, resulting in clean and tidy 3D effects. Obviously, our parallelized encoder can effectively extract both local and global information from the images and accurately capture detailed information by the channel attention module. Thus, our ParaTransCNN method is shown with significant improvement in the segmentation of small organs.

\section{Conclusion and Future Works}
In this paper, we proposed a parallelized encoder based on CNN and Transformer for medical image segmentation. We introduced the pyramid structure in the Transformer to learn global information at different stages. Our ParaTransCNN followed a similar structure as U-Net, where the local and global features were integrated effectively by the channel attention module. We thoroughly compared the segmentation performance of our ParaTransCNN with other state-of-the-art medical image segmentation methods on Multi-Organ and AVT datasets, demonstrating the accuracy of our approach for small organs and vessel segmentation. The proposed ParaTransCNN was shown simple and effective surpassing all state-of-the-art methods by combining the CNN and Transformer in the encoding stage. We believe it can serve as a new benchmark in the field of medical image segmentation. In future work, we plan to continue to explore effective fusion mechanisms between CNN and Transformer to achieve more computer vision and medical image tasks.

\section*{Declaration of Competing Interest}
The authors declare that we have no known competing financial interests or personal relationships that could have appeared to
influence the work reported in this paper.

\section*{Acknowledgments}
The work was partially supported by the National Natural Science Foundation of China (NSFC 12071345). Funding received from the
characteristic \& preponderant discipline of key construction universities in Zhejiang province (Zhejiang Gongshang University-Statistics). Funding was also received from the Collaborative Innovation Center of Statistical Data Engineering Technology \& Application.

\printcredits

\nocite{*}




\end{document}